\documentclass[philosophies,article,accept,pdftex,oneauthor]{Definitions/mdpi}

\newtheorem{Axiom}{Axiom}
\firstpage{1} 
\pubvolume{1}
\issuenum{1}
\articlenumber{0}
\pubyear{2026}
\copyrightyear{2026}
\externaleditor{Firstname Lastname} 
\datereceived{ } 
\daterevised{ } 
\dateaccepted{ } 
\datepublished{ }

\AuthorNames{Gustavo E. Romero}

\Title{A Theory of Partial Truth for Scientific Statements}

\Author{Gustavo 
 E. Romero\orcidA{}}

\address[1]{Instituto Argentino de Radioastronom\'ia (IAR; CCT-La Plata, CONICET; CICPBA; UNLP), Buenos Aires, Argentina;  
 gustavo.esteban.romero@gmail.com 
}

\abstract{Scientific representations often combine accurate and inaccurate components, which a single bivalent classification does not distinguish. This paper develops a restricted formal framework for estimating partial factual truth. For fixed content and valuation criteria, the target is adequacy to the facts, independent of the available evidence; its numerical assessment through evidence remains fallible. Quantitative claims are evaluated by normalized discrepancies between asserted and empirically supported values. Admissible finite conjunctions are assessed over a fixed basis of distinct elementary claims using arithmetic or relevance-weighted means. The resulting valuations preserve the contributions from accurate components and are independent of grouping when component counts or total weights are retained. Zero-relevance additions leave the contextual valuation unchanged, but compensation remains possible among components with positive relevance. Worked examples examine admissibility, changes of granularity, measurement indeterminacy, and sensitivity of comparisons to discrepancy scales and relevance weights. An astrophysical application connects the assessment to published observations. The framework applies to selected testable consequences of models and theories, without providing a numerical calculus for arbitrary formulas or a unique method for choosing scales and weights. It permits conditional comparisons of factual adequacy while distinguishing them from confidence, empirical coverage, and convergence toward a complete truth.
}

\begin{document}

\section{Introduction}
\label{sec:introduction}

Scientific practice often relies on statements, models, and theories that achieve considerable empirical success despite being, strictly speaking, false. Idealizations such as point masses, frictionless surfaces, perfectly rational agents, or infinite populations deliberately misrepresent some features of their targets while accurately capturing others. More generally, a complex scientific representation may combine precise predictions, acceptable approximations, simplifying assumptions, and claims that later prove mistaken. Classifying the entire representation as simply true or false conceals this heterogeneous structure and offers little help in explaining either its empirical success or its contribution to \mbox{scientific progress.}

Classical bivalence assigns each proposition exactly one of two truth values, true or false. The principle of non-contradiction excludes the joint truth of a proposition and its negation in the same respect. Classical logic remains indispensable for establishing consistency, implication, and equivalence. A difficulty arises, however, when binary valuation is used as the sole instrument for assessing the factual adequacy of complex scientific claims. Two false quantitative statements may disagree with the evidence by radically different amounts; two false models may capture different proportions of the relevant behavior of a system; and two theories rejected in their unrestricted form may nevertheless differ greatly in the accuracy and importance of the consequences they retain. These differences invite a more discriminating assessment. The question addressed here is whether factual adequacy can be represented in degrees without confusing truth with probability, vagueness, confirmation, or subjective confidence.

Several established approaches illuminate different aspects of this question. Probabilistic and Bayesian theories assign numerical values to hypotheses in order to represent degrees of confirmation or rational belief in light of evidence \cite{Carnap1950,HowsonUrbach2006}. In Bayesian applications, \(P(p\mid E)\) expresses a probability assigned to the hypothesis \(p\) conditional on the evidence \(E\). It does not ordinarily indicate the degree to which \(p\) itself is true. A hypothesis assigned a high probability within these approaches may nevertheless turn out to be false. Conversely, a proposition known to be approximate may display considerable factual adequacy even when the magnitude of its discrepancy is precisely determined.

On the ontological interpretation {adopted} here, probabilities are attributed to events under specified physical conditions; propositions may state those probabilities but are not themselves bearers of probability. A defense of this ontological position lies beyond the scope of the present paper (see Bunge~\cite{Bunge1988} for a discussion of this issue). The narrower distinction required here is that the proposed valuation \(V_E\) estimates factual adequacy rather than probability: its interpretation and composition rules concern representational discrepancy and the aggregation of elementary content. This distinction can be established without resolving the broader disagreement about the interpretation of probability.

Many-valued and fuzzy logics provide a more direct precedent for assigning values between 0 and 1. Beginning with Zadeh's theory of fuzzy sets and subsequently developed into a family of formal logical systems, fuzzy logic admits graded semantic valuations and defines logical operations over them \cite{Zadeh1965,Hajek1998}. Its most characteristic applications involve vague predicates and gradual classifications: an individual may belong to the class of tall persons, or a temperature to the class of high temperatures, to a certain degree. The resulting values represent degrees of membership or truth associated with predicates whose boundaries are not sharply determined.

The account developed here differs from fuzzy logic in both its target and its formal role. The partial truth of a quantitative scientific statement need not arise from vagueness. A proposition asserting a precise value may have an entirely determinate sense and still agree only approximately with the value indicated by measurement. Its degree of truth then depends on the discrepancy between the asserted and empirically supported values, not on graded membership in a vaguely delimited class. Nor does the proposed valuation replace classical logic with a many-valued calculus. Logical connectives retain their classical meaning, and logical equivalence remains a constraint on valuation. The arithmetic and weighted means introduced below are not truth functions for conjunction; they are aggregation rules for estimating the factual adequacy of the distinct empirical components expressed by a complex proposition. Although fuzzy valuations and the function proposed here both take values in \([0,1]\), their interpretations, domains, and composition rules differ.

Theories of verisimilitude or truthlikeness constitute another important line of research. Initiated by Popper and subsequently developed in more sophisticated forms by Oddie and Niiniluoto, these theories seek to explain how one false theory can be closer to the truth than another \cite{Popper1963,Oddie1986,Niiniluoto1987}. They are particularly relevant to accounts of scientific progress because they permit the comparison of successive theories without requiring the earlier theory to be wholly true. Truthlikeness is commonly formulated as a comparative relation between theories, descriptions, possible worlds, or sets of consequences and a true or maximally accurate description. The question considered in this paper is different. Its primary aim is not to determine the distance between an entire theory and a complete true theory, but to estimate the factual adequacy of empirically evaluable propositions and their components relative to a specified body of evidence. These valuations may subsequently be used to compare models and theories, but they are grounded in testable consequences rather than in direct access to a complete description of the world.

A related approach is provided by the theory of partial structures and the associated notion of pragmatic truth, later called quasi-truth \cite{Mikenberg1986,Bueno1997,daCostaFrench2003}.\endnote{A 
 terminologically related but conceptually distinct account appears in Ingarden's analysis of literary fiction \cite{Ingarden1973} (\S~25). His notion of quasi-judgments concerns sentences that present fictional states of affairs without the ordinary assertive commitment characteristic of factual judgments. This distinction draws attention to the kind of discourse in which a sentence functions. It should be distinguished both from quasi-truth in the theory of partial structures and from the graded factual adequacy considered here, which concerns empirically assessable claims about real systems.} Partial structures represent situations in which the available information does not determine whether certain relations hold for every relevant tuple of objects. A proposition may be regarded as quasi-true when it is true in at least one admissible total extension of such a structure. This framework is well suited to incomplete information, domain-restricted theories, and scientific representations whose applicability remains partly open. Quasi-truth, however, does not generally provide a scalar measure of factual adequacy. It characterizes a form of possible or extendable truth relative to a partial structure rather than assigning a continuous value that measures how well the several components of a proposition agree with the evidence. The approaches may therefore be regarded as complementary. Quasi-truth accommodates the incompleteness and openness of scientific information, whereas the theory developed here seeks to quantify the adequacy supported by the evidence already available.

A further philosophical perspective is offered by Elgin's \textit{True Enough} 
 \cite{Elgin2017}, which examines how scientific models and idealizations can contribute to understanding despite their acknowledged departures from literal truth. Elgin characterizes successful idealizations as ``felicitous falsehoods'': their distortions can facilitate understanding by exemplifying relevant features of the phenomena under investigation. On this account, epistemic value cannot be reduced to representational accuracy, since departures from accuracy may themselves contribute to a representation's epistemic function. The present framework shares the motivation to explain the scientific value of representations that are not wholly true, but addresses a more restricted question: how to estimate the factual adequacy of specified empirical claims relative to a body of evidence. Its relevance weights determine how those claims contribute to the assessment of a particular scientific problem; they do not constitute a measure of understanding or of the epistemic benefits of idealization. Partial factual truth and being ``true enough'' should therefore be distinguished: the former concerns graded factual adequacy under explicit assessment criteria, whereas Elgin's account concerns the contribution of representations to understanding.

An earlier axiomatic line of work treated partial truth as a non-probabilistic semantic magnitude and proposed rules for combining partially true propositions \cite{Bunge2010,Bunge2012,Romero2017,Romero2018}. A subsequent extension distinguished truth from relevance, on the grounds that true but irrelevant statements should not compensate for a false answer to the problem under \mbox{investigation \cite{Romero2017}.} These formulations nevertheless left unresolved questions concerning logical equivalence and the recursive valuation of compound propositions. A recent analysis showed that a binary half-sum rule for conjunction can assign different values to logically equivalent expressions that differ only in their parenthesization \cite{Gracia2025}. This result identifies a formal constraint on any revised account: factual adequacy should not depend on an accidental feature of syntactic grouping.

The objective of this paper is to develop a revised formal theory of partial truth for scientific statements that satisfies this constraint and addresses the preceding issues. Partial factual truth concerns how adequately a proposition represents the facts. For fixed content and valuation criteria, this adequacy is the evidence-independent target of the estimate \(V_E\), rather than its agreement with whichever evidence is currently accepted. Changes in evidence can justify revising the estimate without changing that target; changes in scales, elementary content, or relevance instead change the assessment specification. This distinction does not imply a unique numerical measure independent of all \mbox{valuation criteria.}

For quantitative propositions, the proposed valuation is derived from the normalized discrepancy between the value asserted by the proposition and the value supported by the evidence. This discrepancy is kept distinct from measurement indeterminacy: the latter characterizes the limited determination of measured values, whereas discrepancy characterizes the mismatch between assertion and evidence. Compound propositions are evaluated over sets of distinct elementary components rather than through repeated applications of a binary truth function. Equally relevant components are aggregated by an arithmetic mean, while differences in relevance are represented by a weighted mean. This procedure prevents an isolated falsehood from erasing the adequacy of the remaining components. Relevance weighting excludes contributions from components assigned zero relevance; it does not prevent accurate components with positive relevance from compensating for an important error.

The contribution of this paper is a restricted reformulation of earlier accounts of partial truth, rather than a new mathematical theory of averaging. It combines evidence-relative component valuations, explicit admissibility conditions on elementary content, and relevance-weighted aggregation normalized by total relevance. Logical equivalence is respected within the admitted domain by evaluating the same elementary content, rather than by assigning numerical truth functions to arbitrary compound formulas. The results concerning bounds, endpoints, monotonicity, and grouping are elementary consequences of the aggregation rules. Their role is to establish what the proposed assessment preserves and under which assumptions, not to provide independent claims of mathematical novelty.

The paper begins by distinguishing formal from factual truth and identifying the propositions to which factual valuations can be assigned. It then characterizes scientific bodies of evidence and introduces discrepancy-based valuations for quantitative claims. After examining the limitations of earlier axiomatizations, a revised set of axioms is presented and its principal consequences are derived. Relevance is subsequently incorporated through weighted composition. The discussion examines worked applications and the extension to models and theories, while the conclusion considers the implications for scientific progress. On this account, progress need not be understood as demonstrable convergence toward a final and complete truth. It may instead consist in the construction of representations that become increasingly adequate in the respects most relevant to the problems science seeks to solve.

\section{Formal and Factual Truth}
\label{sec:formal-factual}

The account of partial truth developed here concerns scientific statements that make factual claims. Its scope becomes clearer once factual truth is distinguished from the forms of truth associated with formal systems.

Let \(\mathcal{S}\) be a formal system and \(\varphi\) a formula of its language. Two relations must be distinguished. The expression 

\[\mathcal{S}\vdash\varphi\]

\noindent states that \(\varphi\) is derivable from the axioms of \(\mathcal{S}\) by its rules of inference. This is a syntactic relation. Once the language has been interpreted in a model \(\mathcal{M}\), the expression

\[\mathcal{M}\models\varphi\]

\noindent states that \(\varphi\) is satisfied in \(\mathcal{M}\). This relation is semantic. Derivability depends on the axioms and rules of the system; satisfaction depends on the interpretation assigned to its non-logical terms \cite{Tarski1983}.

Under a fixed classical interpretation, a closed formula receives one of two values:

\[
v_{\mathcal{M}}(\varphi)=
\begin{cases}
1, & \text{if } \mathcal{M}\models\varphi,\\
0, & \text{if } \mathcal{M}\not\models\varphi.
\end{cases}
\]

Formal valuation is therefore bivalent. This does not imply that every formula is decidable within every formal system. If \(\mathcal{S}\) is incomplete, there may be formulas \(\varphi\) such that neither \(\varphi\) nor \(\neg\varphi\) is derivable in \(\mathcal{S}\). Incompleteness concerns the limits of derivability, not the introduction of intermediate semantic values \cite{Godel1931}.

Neither derivability nor satisfaction in an abstract model is sufficient to establish factual truth. A formula may be satisfied in a mathematically admissible model even when that model does not adequately represent any real system. Factual truth concerns a further relation: the adequacy between what a proposition asserts and the facts it \mbox{represents \cite{Bunge1974}.} Its evaluation therefore requires information about the relevant objects, properties, or events---information supplied by observation and experiment.

Let \(\mathcal{P}_{E}\) denote the set of factual propositions open to empirical assessment on the basis of a body of evidence \(E\). The specified valuation criteria determine a domain \(\mathcal{D}_{E}\subseteq\mathcal{P}_{E}\) on which numerical values can be assigned, as detailed in Section~\ref{sec:partial-truth-function}. The corresponding valuation has the form

\[V_{E}:\mathcal{D}_{E}\longrightarrow [0,1],\]

\noindent where \(V_{E}(p)\) represents the degree of factual adequacy attributed to \(p\) with respect to \(E\). The values 1 and 0 represent maximal and zero estimated adequacy on the adopted scale, while intermediate values represent degrees of partial truth. These endpoints do not establish complete correspondence or its complete absence independently of the limitations of \mbox{the assessment.}

The subscript \(E\) records the evidential basis of the attribution; it does not make facts relative to evidence. Whether the state of affairs represented by \(p\) obtains is independent of the procedures used to investigate it. Evidence provides epistemic access to that state of affairs and supplies the grounds on which a value is assigned. When the evidence changes, the attributed value may rationally change even though the proposition and its referents remain the same.

The range \([0,1]\) does not turn \(V_E\) into a many-valued semantics for the object language. Classical logical relations remain in force, and the values assigned by \(V_E\) are not used to redefine negation, conjunction, or implication. The function estimates the factual adequacy of empirically evaluable content. Formal and factual valuations thus serve different purposes: the former concerns satisfaction within formal structures, whereas the latter concerns the adequacy of propositions as representations of reality. Defining the domain of \(V_E\) consequently requires a more precise account of the bearers of factual truth and of the conditions under which they can be empirically evaluated.

\section{Truth Bearers}
\label{sec:truth-bearers}

Truth cannot be attributed indiscriminately to every linguistic, conceptual, or psychological object. Sentences, statements, assertions, judgments, beliefs, and propositions have all been considered possible truth bearers in the philosophical literature \cite{Kunne2003}. Since these entities belong to different ontological categories, the present account requires a precise choice among them.

The term \emph{statement} denotes here a concrete occurrence of a declarative expression: an inscription, utterance, or signal through which something is asserted. Statements are therefore physical objects or events. The same sentence type may have indefinitely many concrete occurrences, and statements produced in different languages or symbolic systems may express the same content. Conversely, expressions with the same syntactic form may express different contents when their interpretation depends on the speaker, time, place, or context of utterance.

Let \(\mathcal{S}_{F}\) be the set of factual statements and \(\mathcal{P}_{F}\) the set of factual propositions. Two statements are \emph{equisignificant} when they have the same reference and express the \mbox{same sense:}

\[
s \sim s'
\quad \Longleftrightarrow \quad
\operatorname{Ref}(s)=\operatorname{Ref}(s')
\ \text{and}\
\operatorname{Sense}(s)=\operatorname{Sense}(s').
\]

A proposition is understood as an equivalence class of equisignificant statements:

\[
[s]=\{s'\in\mathcal{S}_{F}:s'\sim s\}.
\]

The relation \(\sim\) should not be confused with unrestricted lexical synonymy in natural language. Quine objected that cognitive synonymy could not be explained through interchangeability \emph{salva veritate} without appealing to analyticity or necessity and thereby risking circularity. He later argued that the available linguistic and behavioral evidence may fail to determine a unique translation, reference, or meaning \cite{Quine1951,Quine1960}. These objections are directed against attempts to establish a general and independently decidable relation of synonymy on the basis of linguistic dispositions, coextension, or translational \mbox{behavior alone.}

Equisignificance has a narrower function in the present framework. It is attributed only to factual statements considered under a specified interpretation and within a sufficiently explicit scientific and linguistic context. Their reference and sense are fixed by the relevant semantic conventions, theoretical definitions, and intended applications. The relation \(\sim\) is not invoked to define analyticity, nor is it inferred from mere coextension or interchangeability. Two statements may have the same extension or truth value while differing in sense and therefore fail to be equisignificant. When two statements are assigned the same reference and sense under a fixed interpretation, placing them in the same equivalence class records that semantic identity; it does not purport to derive it from linguistic behavior alone. This construction does not settle every Quinean question concerning translation or the determination of meaning, and it does not provide a universal decision procedure for synonymy. It restricts the required identity relation to cases in which the semantic interpretation of the scientific statements has already been made explicit.

With the scope of equisignificance thus delimited, the ontological distinction can be stated more precisely: statements are physical objects, whereas propositions are conceptual objects constituted by equivalence classes of equisignificant statements. Statements written or uttered in different languages may therefore express the same proposition, provided that their reference and sense coincide \cite{Bunge1974a}.

Truth values are attributed primarily to propositions because truth depends on what is asserted, not on the physical properties of the signs used to make the assertion. An inscription may be faint, damaged, or written in blue ink; an utterance may be loud, hesitant, or brief. These properties belong to the statement as a physical object or event. Its truth value, by contrast, derives from the proposition it expresses. Whenever \([s]\in\mathcal{D}_{E}\), the notation \(V_{E}(s)\) may consequently be used as an abbreviation for \(V_{E}([s])\). Hence,

\[
s\sim s'
\quad \Longrightarrow \quad
V_{E}(s)=V_{E}(s').
\]

Several kinds of meaningful expression are excluded from the class of truth bearers. Questions request information, commands prescribe actions, and exclamations express reactions; none of them, considered as such, asserts that a state of affairs obtains. Isolated names and predicates are likewise neither true nor false, and an open formula does not express a complete proposition until values or quantifiers are assigned to its free variables. Such expressions may presuppose, suggest, or be associated with propositions, but they do not themselves belong to the domain of the factual valuation function.

Beliefs, judgments, and opinions are not treated here as primary truth bearers either. They are psychological states or acts and may be rational or irrational, justified or unjustified, sincere or insincere. When a belief is called true or false, the attribution concerns the proposition believed. The same distinction applies to acts of assertion: an assertion may be appropriate, misleading, dishonest, or unwarranted, whereas the proposition asserted is what bears the truth value.

Facts must also be distinguished from truth bearers. A fact does not represent something else and is not made true by its agreement with another entity. Rather, facts are what factual propositions represent and what provide the objective basis for their truth or falsity. This distinction parallels, without presupposing any particular theory of truthmaking, the familiar separation between truth bearers and truthmakers \cite{Mulligan1984}.

Scientific models and theories will not be treated as primitive truth bearers. They are complex representational systems containing mathematical structures, assumptions, idealizations, definitions, and factual claims. Their partial truth must therefore be evaluated derivatively, through finite and representative sets of factual propositions expressing their empirically testable consequences. This extension will be considered after the valuation of individual propositions has been established.

Finally, not every factual proposition belongs to the domain of \(V_{E}\). A proposition must possess a sufficiently precise sense, an identifiable reference, and consequences that can be assessed in relation to the available evidence. An expression whose sense or reference cannot be determined does not specify an assessable factual proposition. A proposition for which no relevant evidential assessment is possible likewise lies outside the valuation domain and should not be assigned the value zero. Thus,

\[
\mathcal{P}_{E}\subseteq\mathcal{P}_{F},
\]

\noindent where \(\mathcal{P}_{E}\) contains those factual propositions that are meaningful and open to empirical assessment with respect to the body of evidence \(E\). These are necessary conditions for membership in \(\mathcal{D}_{E}\); the additional valuation criteria determine whether the numerical procedure applies. Specifying the nature and structure of such evidence is therefore the next task.

\section{Truth and Evidence}
\label{sec:truth-evidence}

Let \(p\) be a factual proposition and let \(E\) be a body of evidence relevant to its evaluation. We call a \emph{body of evidence} a structured set of basic statements reporting the results of scientifically controlled observational or experimental procedures.

A basic statement records the result of an observation, measurement, detection, or experiment, together with the relevant conditions and associated measurement indeterminacy. It is basic not because it is infallible or free of theoretical presuppositions, but because it functions as an empirical premise in the evaluation at hand.\endnote{Neurath's protocol sentences were themselves revisable statements formulated in a public physicalist language, not infallible reports of private experience \cite{Neurath1932}. Our use of ``basic statement'' shares this rejection of an incorrigible empirical foundation, but does not prescribe Neurath's particular linguistic form, with its explicit references to the observer and acts of perception. A statement is basic here by virtue of its methodological role in a given evaluation. Its acceptance requires empirical controls; consistency with other accepted statements is not, by itself, sufficient to establish factual adequacy.} Its acceptance remains provisional and open to further testing \cite{Popper1959} (Sections~28--30), 
 and requires documented procedures, appropriate methodological controls, and results open to independent scrutiny. A collection of assertions does not constitute a body of scientific evidence merely because those assertions have been recorded or widely accepted: their provenance, reliability, and relevance must be assessed.

The attribution of a truth value is expressed by a metalinguistic proposition:
\begin{quote}
``On the basis of the evidence \(E\), the proposition \(p\) is assigned the truth value \(V_{E}(p)\).''
\end{quote}

The 
 subscript \(E\) makes explicit the evidential basis of the valuation. The attributed value expresses an estimate of how adequately \(p\) represents the facts, rather than a degree of confidence in its truth.

Factual truth is not reducible to evidence. Whether things are as \(p\) asserts does not depend on our acceptance of the proposition or on the evidence available to us. Evidence provides the empirical basis for estimating its factual adequacy; it does not constitute the facts against which that adequacy is assessed. An attribution of truth can therefore be justified by the available evidence and still require subsequent revision.

With the meaning and reference of \(p\) held fixed, replacing \(E\) with a different body of evidence \(E'\) may yield a different attributed value, \(V_{E'}(p)\). Such a difference is possible, not necessary: additional evidence may leave the valuation unchanged. When a revision is warranted, it is our assessment of the proposition that changes. A revision of our knowledge does not, by itself, alter the facts described. Holding the valuation criteria fixed allows us to distinguish changes due to the evidence from changes due to the procedure used to evaluate it.

Meaning precedes testing. To test a proposition, we must understand what it asserts, what it refers to, and what observational or experimental results would bear on its evaluation. Yet understanding a proposition does not require knowing its truth value: we can understand an assertion whose factual adequacy remains undetermined. Truth presupposes meaning, whereas understanding meaning does not presuppose knowledge of truth.

\section{Quantitative Propositions}
\label{sec:quantitative}

For quantitative propositions, comparison with evidence begins with the values asserted and measured. Consider a proposition \(p\) that assigns a value \(\mu\) to a scalar property \(M\) of a specified object or system:
\[
p:\quad M = \mu.
\]
Suppose now 
 that an empirical procedure yields an estimate of the same property:
\[
M = e \pm \delta,
\]
where \(e\) is the measured value and \(\delta > 0\) is the half-width of the reported measurement interval. The term \emph{measurement indeterminacy} can be used to denote the limited determination of the value of \(M\) by the available procedures and evidence. It designates neither a psychological state of the investigator nor an intrinsic lack of a definite value in the measured system. The convention used to construct the interval and, where applicable, its coverage level must be specified \cite{JCGM2008}. Then the basic statement recording the result belongs to \(E\).

A simple criterion of compatibility is that the asserted value \(\mu\) lies within the reported interval:
\[
|\mu - e| \leq \delta.
\]
For example, with all values expressed in the same units, let \(p\) assert that \(\mu = 9.80\), and suppose that the measurement gives
\[
M = 9.79 \pm 0.02.
\]
Since
\[
|9.80 - 9.79| = 0.01 < 0.02,
\]
then the asserted value is compatible with the measurement under this criterion.

This criterion determines whether an asserted value lies inside or outside the interval, but does not yet assign a degree of truth. Two propositions may satisfy it while differing in their distance from the central estimate \(e\). Such differences do not necessarily imply that the measurement can discriminate between the asserted values, but they can be represented by a measure of discrepancy.

Measurement indeterminacy and discrepancy play different roles. The former concerns the values attributable to \(M\) on the basis of the measurement; the latter measures the departure of the asserted value \(\mu\) from the empirical estimate \(e\). For a scalar quantity, an elementary measure of this discrepancy is
\[
d_{E}(p) = |\mu - e|.
\]

To compare discrepancies across different quantities or scales, we need a dimensionless measure. Let \(\beta_{E}(p)\) denote the normalized relative discrepancy of \(p\) with respect to \(E\), with
\[
0 \leq \beta_{E}(p) \leq 1.
\]
Given a characteristic scale \(s_{E}(M) > 0\), appropriate to the problem and expressed in the same units as \(\mu\) and \(e\), we may define
\[
\beta_{E}(p) =
\min\left\{1,\frac{|\mu - e|}{s_{E}(M)}\right\}.
\]
The normalization rule must be specified and applied consistently to the propositions being compared.

When \(M\) has a non-arbitrary zero and \(e \neq 0\), a simple choice is \(s_{E}(M) = |e|\), provided that the measured value supplies an appropriate reference scale. If \(e = 0\), or if the origin of the measurement scale is arbitrary, another normalization is required. Extending the construction to vector quantities or distributions also requires an appropriate measure of discrepancy in place of the scalar absolute difference.

The characteristic scale specifies the magnitude of discrepancy against which departures from the empirical estimate are assessed. It must have an interpretation independent of the desired valuation, such as the magnitude of a nonzero reference value on a ratio scale or a scientifically specified range of variation. Dividing by this scale makes the discrepancy invariant under a consistent change of measurement units, but does not by itself establish comparability across different properties or assessment purposes.

Truncation introduces an explicit resolution limit: discrepancies at or above the characteristic scale receive the same maximal normalized discrepancy. Under the valuation \(V_E(p)=1-\beta_E(p)\), they consequently receive zero. This endpoint means that the adopted rule assigns no remaining adequacy credit to the quantitative assertion; it does not mean that all larger discrepancies are empirically equivalent. The untruncated discrepancy should therefore remain available whenever differences beyond this threshold matter.

The linear complement adopted in Axiom 4 gives equal reductions in valuation to equal increases in normalized discrepancy below the truncation threshold. Together with the endpoint assignments, this constant-rate assumption determines the linear mapping. It is a substantive calibration choice.

The choice also affects comparisons between compound representations. Consider two descriptions with equally relevant components and normalized discrepancy profiles \(\beta_A=(0,1)\) and \(\beta_B=(0.6,0.6)\). Applying \(V=1-\beta\) and then averaging gives scores of \(0.50\) and \(0.40\), respectively. The alternative component rule \(V'=1-\beta^2\), which also preserves the endpoints and decreases monotonically with discrepancy, instead gives aggregate scores of \(0.50\) and \(0.64\). Thus, preserving the ordering of individual discrepancies does not preserve aggregate rankings under a nonlinear transformation.

Unlike the linear rule, the quadratic penalty makes the loss associated with a fixed increment depend on the initial discrepancy: its marginal rate increases with \(\beta\). This does not mean that the quadratic penalty is everywhere larger than the linear penalty; for \(0<\beta<1\), \(\beta^2<\beta\), so the alternative assigns a higher component score. These are different assessment conventions, and the correspondence account alone does not select between them. The linear rule therefore expresses a restricted modeling assumption, not a uniquely correct numerical measure of partial truth. Where the scientific purpose supports alternative penalty functions, their effects on the comparison should be examined alongside sensitivity to scales and relevance weights.

Under the adopted linear rule, a higher value indicates a smaller normalized discrepancy, with strict ordering lost where truncation produces ties. Numerical differences express differences on the adopted linear adequacy scale. A valuation of \(0.8\) does not establish that a proposition is ``twice as true'' as one valued at \(0.4\). Comparisons across properties require a further justification that their scales express comparable departures; mapping all values into \([0,1]\) is insufficient.

Thus, \(\beta_{E}(p)\) quantifies disagreement with the empirical estimate relative to a specified scale, whereas \(\delta\) characterizes the measurement indeterminacy associated with \mbox{that estimate.}

\section{Bodies of Evidence}
\label{sec:bodies-evidence}

Scientific evaluations commonly draw on several observations or measurements. A finite body of evidence can be represented as
\[
E = \{\eta_{1},\eta_{2},\ldots,\eta_{n}\},
\]
where each \(\eta_{i}\) records the result of a documented and scientifically controlled observational or experimental procedure. Each statement, together with its supporting documentation, should identify the relevant system, the property investigated, the procedure and conditions of measurement, and the result with its associated measurement indeterminacy. The elements of \(E\) may differ in precision and reliability and need not be independent.

Combining measurements of the same quantity under comparable conditions requires an explicit treatment of their measurement indeterminacy and any correlations arising from shared instruments, calibration procedures, or data \cite{JCGM2008} (Section~5.2). Agreement between results does not eliminate the possibility of a shared systematic error. When the results are compatible and their joint analysis is justified, they may be summarized by an \mbox{effective estimate:}
\[
M = e_{E} \pm \delta_{E},
\]
where \(e_{E}\) is the value inferred from the relevant measurements in \(E\), and \(\delta_{E}\) is the half-width of the resulting measurement interval under the stated convention for constructing the interval. When such a summary is adequate, \(e_{E}\) replaces the individual estimate \(e\) in the discrepancy measure introduced in Section~\ref{sec:quantitative}.

Under comparable conditions and the same convention for constructing the intervals, a smaller \(\delta_{E}\) permits a more stringent test of an asserted value. The width of a reported interval does not, by itself, establish the quality of the evidence. Faulty calibration, unaccounted systematic effects, or an inadequate measurement model may require a basic statement to be revised or withdrawn. The assessment must therefore consider how the results were obtained and how their measurement intervals were established.

If two determinations are incompatible, the one reporting the narrower measurement interval should not automatically be preferred. Their disagreement calls for an examination of the measurement procedures, experimental conditions, and assumptions used in the analysis. Any unresolved discrepancy must remain explicit in the assessment of \(E\) and in the limitations of the resulting valuation.

The valuation \(V_{E}\) thus rests on a joint assessment of the relevant results, their dependencies, and their methodological support. An isolated measurement cannot be evaluated without regard to the other relevant evidence.

\section{The Partial-Truth Function}
\label{sec:partial-truth-function}

Complete factual truth would require full correspondence between what a proposition asserts and the facts it describes. The available evidence provides the basis for estimating that correspondence, but its finite and revisable character limits what any particular valuation can establish.

For a specified body of evidence \(E\) and a fixed set of valuation criteria, we introduce the function
\[
V_{E}:\mathcal{D}_{E} \rightarrow [0,1],
\]
where \(\mathcal{D}_{E} \subseteq \mathcal{P}_{E} \subseteq \mathcal{P}_{F}\). Here \(\mathcal{P}_{E}\) comprises meaningful factual propositions open to empirical assessment through \(E\), whereas \(\mathcal{D}_{E}\) contains those to which the specified valuation criteria permit a numerical assignment. Fixing those criteria may therefore narrow the domain; empirical assessability alone does not guarantee that the present numerical procedure applies. For each \(p \in \mathcal{D}_{E}\), the value \(V_{E}(p)\) represents its estimated degree of \mbox{factual adequacy.}

The elementary basis \(\mathcal{B}_{E}\subseteq\mathcal{D}_{E}\), introduced in Section~\ref{sec:elementary}, selects the claims treated as undecomposed components at the adopted level of analysis. It is not the entire valuation domain, which also includes admissible compound propositions.

The interpretation of this valuation rests on the correspondence account adopted here: factual propositions represent determinate features of real systems, and their adequacy depends on how accurately they represent those features. For fixed propositional content and valuation criteria, adequacy to the facts is the evidence-independent target that \(V_E\) seeks to estimate. In the scalar case, this target can be understood by replacing the empirical estimate in the discrepancy rule with the actual value of the property, while holding the characteristic scale fixed. This conceptual distinction does not provide access to that actual value or select a unique scale. It separates the target of assessment from the empirical information used to estimate it.

For a quantitative proposition, the asserted value and the empirical estimate must refer to the same property, system, and conditions. Agreement with the estimate supports an attribution of factual adequacy only insofar as the empirical procedure is independently warranted as a reliable determination of that property. Relevant grounds include calibration against appropriate standards, examination of systematic effects and measurement-model assumptions, and comparison with independent determinations where available. Mere agreement with an accepted statement or a published number is insufficient. These methodological grounds, discussed in Section~\ref{sec:bodies-evidence}, justify the inferential step from agreement with evidence to estimated adequacy to facts without guaranteeing its correctness.

A biased measurement illustrates the distinction. Suppose, hypothetically, that a property's actual value is \(100\) units, but an undetected calibration bias produces an estimate of \(110\) units. Compare propositions asserting \(100\) and \(110\) units, using a fixed characteristic scale of \(100\) units. Their evidence-relative valuations are \(0.9\) and \(1\), respectively, although their adequacy values against the stipulated actual value, under the same criterion, are \(1\) and \(0.9\). The procedure therefore favors the less accurate claim when its empirical reference is biased. If subsequent calibration identifies and corrects the bias, an estimate of \(100\) units reverses the assigned scores. The target remains unchanged; the empirical basis of its estimation has improved.

If a detected systematic problem cannot be corrected or bounded sufficiently to support the assessment, the affected valuation should be withheld. If defensible bounds remain available but permit alternative rankings, the central scores may be reported with those limits, while the comparison remains unresolved, as illustrated in Section~\ref{sec:rival-comparison}. Thus, the numerical construction is operationally a measure of agreement with an empirical estimate, but its interpretation as estimated partial factual truth depends on the methodological warrant connecting that estimate to the represented property. Neither the formula nor the adoption of a correspondence account alone supplies that warrant.

The restriction to \(\mathcal{D}_{E}\) makes the valuation generally partial over \(\mathcal{P}_{F}\): some factual propositions may remain unevaluated. This restriction concerns whether a value can be assigned, whereas partial truth concerns the degree of adequacy represented by an \mbox{assigned value.}

The upper endpoint,
\[
V_{E}(p) = 1,
\]
represents the maximum estimated adequacy under the adopted valuation criteria. It does not imply that complete correspondence with the facts has been conclusively established.

The lower endpoint,
\[
V_{E}(p) = 0,
\]
represents zero estimated adequacy on the adopted scale. It must not be confused with the absence of sufficient evidence to evaluate \(p\): when \(p \notin \mathcal{D}_{E}\), its valuation is undefined, \mbox{not zero.}

Intermediate values,
\[
0 < V_{E}(p) < 1,
\]
express degrees of partial truth: the proposition is attributed some factual adequacy without attaining the maximum of the adopted scale. Both the intermediate values and the endpoints remain relative to the specified evidence and valuation criteria.

\section{Formal Structure and Interpretation}
\label{sec:formal-structure}

Axioms constrain the relations among the elements of a mathematical theory without, by themselves, specifying what those elements represent. The same formal structure may therefore admit different empirical interpretations. This distinction is explicit in Kolmogorov's presentation of probability theory, which separates its abstract axiomatic formulation from its concrete interpretations \cite{Kolmogorov1950} (Chapter~I).

The continuity equation, 
\[
\frac{\partial\rho}{\partial t}
+ \nabla \cdot \mathbf{J} = 0,
\]
illustrates this distinction. Here \(\rho\) denotes a density and \(\mathbf{J}\) the corresponding flux. Depending on the quantities represented, the equation may express local conservation of mass in a fluid, electric charge, or, in its one-dimensional form, the number of vehicles along a road. In each application, the absence of a source term requires that the relevant quantity be locally conserved. The mathematical relation remains the same, while the meanings of its symbols and the systems to which it applies differ.

Likewise, the use of the interval \([0,1]\) does not determine what a valuation measures. For example,
\[
V_{E}(p) = 0.8
\]
means that \(p\) has an estimated degree of factual adequacy of \(0.8\), relative to \(E\) and the adopted valuation criteria. This is a valuation of its factual adequacy, not a probability attributed to \(p\). If \(p\) itself asserts an event probability, that probability must be distinguished from the valuation of \(p\).

The distinction between interpretation and formal structure also requires care when drawing analogies with probability theory. Kolmogorov's framework imposes specific structural conditions, including additivity over mutually exclusive events. Sharing a numerical scale does not entail satisfying those conditions, and changing the interpretation of a function does not by itself alter its mathematical properties.

The rules for \(V_{E}\) must therefore be specified and justified in terms of the factual adequacy they are intended to represent. In particular, the composition rules for compound propositions concern how the adequacy of their components contributes to the valuation of the whole. They are not rules for calculating probabilities of events.

\section{A Difficulty with Earlier Formulations}

\label{sec:earlier-formulations}

Earlier formulations of partial truth proposed general rules for the negation, conjunction, and disjunction of partially true propositions \cite{Bunge2010,Bunge2012}. Their subsequent axiomatic development retained formal difficulties that bear directly on the interpretation of the valuation function.

For conjunctions whose components are not negations of one another, later formulations adopted the rule \cite{Romero2017,Romero2018}

\[
V_{E}(p \land q)
=
\frac{V_{E}(p) + V_{E}(q)}{2}.
\]

When applied recursively, this rule is incompatible with the requirement that logically equivalent propositions receive the same value. Gracia Di Rienzo \cite{Gracia2025} establishes this incompatibility through the associativity of conjunction:

\[
(p \land q) \land r
\equiv
p \land (q \land r).
\]

Following his counterexample, let

\[
V_{E}(p) = 1,
\qquad
V_{E}(q) = 0.5,
\qquad
V_{E}(r) = 0.
\]

Grouping the first two components gives

\[
V_{E}\left((p \land q) \land r\right)
=
\frac{\frac{1 + 0.5}{2} + 0}{2}
=
0.375,
\]

\noindent whereas grouping the last two gives

\[
V_{E}\left(p \land (q \land r)\right)
=
\frac{1 + \frac{0.5 + 0}{2}}{2}
=
0.625.
\]

The two equivalent expressions therefore receive different values. The source of the discrepancy is the recursive use of the binary arithmetic mean. In the first grouping, \(p\), \(q\), and \(r\) receive weights \(1/4\), \(1/4\), and \(1/2\), respectively; in the second, their weights are \(1/2\), \(1/4\), and \(1/4\). A change in parentheses thus changes the contribution of each component without changing what the conjunction asserts.

The finite arithmetic mean itself is not introduced here. The earlier formulations already proposed averaging the values of a finite collection of conjuncts and distinguished truth from relevance \cite{Romero2017,Romero2018}. The 2018 formulation also considered a relevance-adjusted average, dividing the sum of relevance--truth products by the number of components \cite{Romero2018}. The present contextual valuation instead divides that sum by total relevance, provided that this total is positive. This difference has a direct consequence: adding a component with zero relevance leaves the valuation unchanged. For example, a component with valuation \(0.8\) and relevance \(1\), combined with a component of relevance \(0\), yields \(0.4\) when the denominator counts both components, but \(0.8\) when it is total relevance. The latter normalization is adopted here to assess the adequacy of the content that contributes to the specified problem.

The principal change in logical scope is equally important. The earlier accounts proposed general numerical rules for logical connectives, whereas the present account evaluates admissible finite conjunctions through a fixed set of distinct elementary components. Gracia Di Rienzo's criticism concerns the compatibility of the earlier valuation rules with logical equivalence, with the associativity counterexample providing a particularly transparent instance \cite{Gracia2025}. The present reformulation addresses that instance by abandoning recursive binary averaging and retaining component counts and weights correctly when combining disjoint groups. It does not establish a consistent numerical calculus for unrestricted negation and disjunction. Its grouping results are conditional on the adopted elementary basis and do not establish invariance under changes of that basis.

A finite conjunction within the evaluable domain is assessed from its full, specified decomposition into elementary components. When these components have equal weight, their values are averaged together. The half-sum is consequently the special case of two equally weighted elementary components, not a general rule for combining the values of arbitrary compound expressions. In the preceding example, either grouping is evaluated from the same three components and receives the value

\[
\frac{1 + 0.5 + 0}{3} = 0.5.
\]

This removes the dependence on parenthesization for a fixed elementary decomposition. It does not, by itself, establish invariance under every classical equivalence; that requirement also constrains the identification of elementary components and the treatment of logically redundant content.

\section{Axioms of Partial Factual Truth}

\label{sec:axioms}

The following axioms govern elementary valuations and aggregation over admissible finite sets of elementary content. They do not define a recursive valuation for arbitrary formulas; the logical connectives retain their classical interpretation.

\subsection{Elementary Content and Independence}

\label{sec:elementary}

Let \(\mathcal{B}_{E} \subseteq \mathcal{D}_{E}\) be a fixed basis of elementary factual propositions evaluable with respect to the evidence \(E\). A proposition is elementary at the adopted level of analysis when it is evaluated as a single factual claim rather than decomposed into further components. The choice of basis forms part of the valuation criteria and must be held fixed when \mbox{comparing representations.}

For the purposes of composition, a finite set of elementary propositions is independent when none of its members is a logical consequence of the remaining members and their empirical contents do not merely repeat the same assertion. We also require joint consistency: the components must not collectively entail a contradiction. Independence and consistency are assessed under a fixed interpretation and background theory.

These conditions concern the content of the propositions, not the statistical independence of the measurements used to evaluate them. Repeated measurements bearing on the same assertion contribute to the body of evidence; they do not create additional components of the proposition. Correlations arising from shared instruments, calibration procedures, or data must be taken into account in the joint assessment of \(E\) \cite{JCGM2008}.

The composition rule applies only when a proposition \(P\) determines a unique finite, nonempty subset \(S(P)\) of the chosen basis whose members are distinct, jointly consistent, and independent, and whose conjunction is logically equivalent to \(P\):

\[
S(P) = \{p_{1},p_{2},\ldots,p_{n}\}
\subseteq \mathcal{B}_{E},
\qquad
P \equiv \bigwedge_{i=1}^{n}p_{i}.
\]

Uniqueness is an additional condition of admissibility, not a consequence of independence alone. If two distinct subsets satisfy the component conditions and their conjunctions represent the same proposition, the composition rule cannot be applied until the ambiguity has been resolved.

For propositions satisfying this condition, valuation applies to \(S(P)\), not to the syntactic presentation of \(P\). Reordering, regrouping, or repeating components does not change this set. In particular,

\[
p \land p \equiv p,
\]

\noindent so repeating \(p\) introduces no additional component. More generally, logically equivalent expressions are evaluated through the same \(S(P)\) and receive the same value. This invariance holds relative to the fixed basis; it does not establish invariance under changes of basis.

\subsection{Valuation Axioms}

For a fixed body of evidence and specified valuation criteria, we adopt the following axioms. The first two state the bounds and interpretation of the scale; the remaining two specify composition and quantitative valuation.

The composition axiom applies only to the admissible finite conjunctions specified above. The domain \(\mathcal{D}_{E}\) is not assumed to be closed under negation or disjunction, and no general numerical rules for these connectives are postulated. Expressions involving them may nevertheless receive a valuation when they represent propositions already admitted to the domain. Their values are then determined by the adopted empirical criteria and, where applicable, their elementary component sets, rather than by a recursive evaluation of connectives.

\begin{Axiom}[Boundedness
]
For every \(p \in \mathcal{D}_{E}\),
\[
0 \leq V_{E}(p) \leq 1.
\]
\end{Axiom}

\begin{Axiom}[Normalization]

The maximum degree of factual adequacy supported by the evidence under the adopted criteria is assigned the value
\[
V_{E}(p) = 1,
\]whereas the minimum degree on the adopted scale is assigned the value
\[
V_{E}(p) = 0.
\]

\end{Axiom}

These endpoints concern estimated factual adequacy. A value of 1 does not establish complete correspondence with the facts independently of the limitations of the evidence, and a value of 0 does not represent the absence of sufficient evidence for evaluation.

\begin{Axiom}[Composition]
Let \(P\) satisfy the conditions of the preceding subsection, with

\[
S(P) = \{p_{1},p_{2},\ldots,p_{n}\}.
\]

If its elementary components have equal positive relevance to the problem under consideration, then

\[
V_{E}(P)
=
\frac{1}{n}
\sum_{i=1}^{n}V_{E}(p_{i}).
\]
\end{Axiom}

For an admissible conjunction with two equally relevant components, \(S(P)=\{p,q\}\), this reduces to the half-sum:

\[
V_{E}(p \land q)
=
\frac{V_{E}(p) + V_{E}(q)}{2}.
\]

The half-sum does not apply recursively to arbitrary compound subexpressions. The valuation of the whole conjunction is calculated from its complete elementary \mbox{component set.}

A multiplicative aggregation, instead, would assign zero to the entire conjunction whenever any component received zero:

\[
\prod_{i=1}^{n}V_{E}(p_{i}) = 0.
\]
This property clearly makes multiplicative aggregation an inadequate choice for composition. 

The arithmetic mean, on the contrary, preserves the contribution of the remaining components. This property alone, however, does not determine the aggregation rule; its assumptions of comparability and compensation, and an alternative based on the minimum, are discussed in Section~\ref{sec:relevance-content}.

When components differ in relevance, the arithmetic mean is replaced by the weighted expression introduced in Section~\ref{sec:weighted-composition}.

\begin{Axiom}[Quantitative valuation]

If \(p \in \mathcal{B}_{E}\) is a quantitative proposition evaluated against the evidence \(E\), and \(\beta_{E}(p)\) is its normalized discrepancy as specified in Section~\ref{sec:quantitative}, then

\[
V_{E}(p) = 1 - \beta_{E}(p),
\qquad
0 \leq \beta_{E}(p) \leq 1.
\]\end{Axiom}

When \(\beta_{E}(p) = 0\), the asserted value coincides with the empirical estimate and receives the maximum valuation on the adopted scale. This agreement does not eliminate measurement indeterminacy.

When \(\beta_{E}(p) = 1\), the valuation is zero. For the truncated normalization introduced earlier, this occurs when the discrepancy reaches or exceeds the characteristic scale \(s_{E}(M)\); it need not represent the largest discrepancy physically possible.

\section{Immediate Consequences}

\label{sec:immediate-consequences}

Throughout this section, all conjunctions satisfy the admissibility conditions of \mbox{Section~\ref{sec:elementary}.} Their elementary components belong to the same fixed basis and are distinct, jointly consistent, independent, and equally relevant. Let

\[
P = \bigwedge_{i=1}^{n}p_{i},
\qquad n \geq 1,
\]

\noindent with

\[
S(P) = \{p_{1},p_{2},\ldots,p_{n}\}.
\]

The endpoint values are understood in the sense specified by Axiom 2.

\begin{enumerate}[label=,leftmargin=0.01em,labelsep=4mm]
\item\textit{Elementary properties.} 

Axioms 1 and 3 immediately give
\end{enumerate}
\[
0\leq\min_i V_E(p_i)\leq V_E(P)\leq\max_i V_E(p_i)\leq1.
\]
The mean equals 1 if and only if every component value is 1, and equals 0 if and only if every component value is 0. These statements follow because all component values lie in \([0,1]\): any value below 1 lowers the mean below 1, and any positive value raises it above 0. Increasing a component value by \(\Delta\), within the permitted range and with all other values fixed, increases the mean by \(\Delta/n\).

If one component has value 0 and the remaining \(n-1\) have value 1, then \(V_E(P)=1-1/n\). For example, nine unit values and one zero give \(V_E(P)=0.9\). The zero lowers the aggregate without canceling the adequacy of the other components; this is not the classical truth table for conjunction.

For quantitative components evaluated by Axiom 4, substitution into the mean gives
\[
V_E(P)=1-\bar\beta_E(P),\qquad
\bar\beta_E(P)=\frac{1}{n}\sum_{i=1}^{n}\beta_E(p_i).
\]
Thus, aggregate adequacy is one minus the mean normalized discrepancy. These properties support the interpretation of the rule; the following proposition specifies how disjoint groups can be combined without changing their aggregate valuation.

\begin{Proposition}[Grouping invariance]
Let \(P\) and \(Q\) be admissible conjunctions with disjoint elementary component sets containing \(m\) and \(n\) members, respectively. Suppose that \(P\land Q\) is also admissible, with \(S(P\land Q)=S(P)\cup S(Q)\) and equal relevance throughout. Then
\[
V_{E}(P\land Q)
=
\frac{m\,V_{E}(P)+n\,V_{E}(Q)}{m+n}.
\]
\end{Proposition}

\begin{proof}
Since 
 the component sets are disjoint, Axiom 3 gives

\[
V_{E}(P\land Q)
=
\frac{
\displaystyle\sum_{p\in S(P)}V_{E}(p)
+
\displaystyle\sum_{q\in S(Q)}V_{E}(q)
}{m+n}.
\]

The two sums are \(m\,V_{E}(P)\) and \(n\,V_{E}(Q)\), respectively, which yields the stated expression.

For three groups \(P\), \(Q\), and \(R\), containing \(m\), \(n\), and \(\ell\) elementary components, assume that their component sets are pairwise disjoint and that all conjunctions involved remain admissible. Either grouping then yields

\[
\frac{
m\,V_{E}(P)+n\,V_{E}(Q)+\ell\,V_{E}(R)
}{m+n+\ell}.
\]

Consequently,

\[
V_{E}\left((P\land Q)\land R\right)
=
V_{E}\left(P\land(Q\land R)\right).
\]

The result depends on retaining the number of elementary components in each group. It does not make the unweighted binary arithmetic mean associative.

\end{proof}

\begin{Example}[Grouping and logical equivalence]

Consider 
 five elementary propositions satisfying the preceding conditions, with

\[
V_{E}(p_{1})=1,
\qquad
V_{E}(p_{2})=0.8,
\qquad
V_{E}(p_{3})=0.6,
\]

\noindent and

\[
V_{E}(q_{1})=0.4,
\qquad
V_{E}(q_{2})=0.2.
\]
\end{Example}

Let

\[
P=p_{1}\land p_{2}\land p_{3},
\qquad
Q=q_{1}\land q_{2}.
\]

Their valuations are

\[
V_{E}(P)=\frac{1+0.8+0.6}{3}=0.8,
\qquad
V_{E}(Q)=\frac{0.4+0.2}{2}=0.3.
\]

Now consider

\[
F=\neg(\neg P\lor\neg Q).
\]

By classical logical equivalence, \(F\equiv P\land Q\), so its elementary component set is

\[
S(F)=\{p_{1},p_{2},p_{3},q_{1},q_{2}\}.
\]

Proposition 1 therefore gives

\[
V_{E}(F)
=
\frac{3(0.8)+2(0.3)}{5}
=
0.6.
\]

Taking the unweighted half-sum of \(V_{E}(P)\) and \(V_{E}(Q)\) would instead give \(0.55\), because it would assign equal weight to groups containing different numbers of elementary components.

The expression \(F\) is evaluated by identifying its elementary content through classical equivalence. No intermediate numerical valuations of \(\neg P\), \(\neg Q\), or their disjunction are required.

\begin{enumerate}[label=,leftmargin=0.01em,labelsep=4mm]
\item \textit{Relation to probability.} 

For two events \(A\) and \(B\) in the same probability space, the probability of their joint occurrence satisfies

\[
\Pr(A\cap B)
\leq
\min\{\Pr(A),\Pr(B)\},
\]whereas the arithmetic mean in Axiom 3 exceeds the smaller component value whenever the two values differ. The reference to Kolmogorov concerns an axiomatic approach to numerical valuation, not the adoption of the probability axioms \cite{Kolmogorov1950}.
\end{enumerate}

\section{Truth and Relevance}

\label{sec:truth-relevance}

\subsection{The Problem of Relevance}

\label{sec:problem-relevance}

Axiom 3 concerns components of equal relevance. Scientific descriptions may also contain claims that contribute unequally to answering a particular question. Extending the valuation to such descriptions requires an explicit account of relevance, a requirement already recognized in earlier work \cite{Romero2017}.

Consider a hypothetical laboratory assessment of the electrical resistance \(R\) of a specimen at a specified temperature. Let \(F\) be the problem of assessing a prediction of this resistance against a direct electrical measurement. Suppose that the evidence gives

\[
R=(100.0\pm1.0)\,\Omega,
\]

\noindent whereas the proposition under evaluation asserts

\[
p_{0}:\quad R=10.0\,\Omega.
\]

With the characteristic scale set to the empirical estimate, \(s_E(R)=100.0\,\Omega\), \mbox{Axiom 4 gives}

\[
V_E(p_{0})
=1-\frac{|10.0-100.0|}{100.0}
=0.1.
\]

Suppose that the laboratory report also contains \(n\) propositions \(p_i\) about distinct geometrical quantities associated with fiducial marks on the specimen holder. Each asserts the central value obtained in a separate optical inspection and therefore receives \(V_E(p_i)=1\) under the adopted discrepancy criterion. Assume that all these claims satisfy the structural admissibility conditions and that the marks play no role in the electrical measurement model, calibration, or corrections.

An indiscriminate unweighted average would give

\[
\frac{V_E(p_{0})+\sum_{i=1}^{n}V_E(p_i)}{n+1}
=\frac{0.1+n}{n+1}.
\]

For nine such additional claims, this average is \(0.91\), although the resistance prediction has not improved. The calculation falls outside the equal-relevance condition of Axiom 3. It illustrates how a high average adequacy across an extended description can be misleading if presented as the adequacy of its answer to a particular question.

\subsection{Relevance to a Scientific Problem}

\label{sec:relevance-problem}

Relevance depends on the problem and its scientific context. The geometrical claims in the example do not contribute to assessing the resistance prediction, but they may be central to checking the specifications of the specimen holder. Sharing a reference or experimental setting does not guarantee equal relevance.

The present proposal concerns how the importance of a proposition's content for a specified scientific problem enters the assessment of factual adequacy. The relevance weights introduced below require criteria appropriate to the investigation; the composition rule specifies how they enter the contextual valuation. Section~\ref{sec:relevance-content} compares this role with Schurz's accounts of deductive relevance and content elements.

For a specified problem \(F\), we introduce a relevance function on factual propositions:

\[
\operatorname{Rel}_F:\mathcal{P}_F\longrightarrow[0,1].
\]

The value 0 indicates that the content of a proposition does not bear on the question under the adopted assessment criteria; 1 indicates maximal relevance. Intermediate values express differences in importance. Restricting the range to \(\{0,1\}\) gives the special case in which only relevant and irrelevant claims need to be distinguished.

Relevance concerns the role of a proposition's content in answering the question, independently of whether that content is accurate. An incorrect prediction of the target resistance is directly relevant to its assessment. Conversely, an accurate claim about an unrelated geometrical feature remains accurate even when it has no relevance to \(F\). Thus \(V_E(p)\) estimates factual adequacy, whereas \(\operatorname{Rel}_F(p)\) specifies the importance of that content for the problem.

No universal assignment of relevance values is assumed. Each application must state and justify its criterion, using the scientific objective and the role of the propositions in addressing it. A precise numerical claim is not automatically relevant, and a qualitative, causal, or structural claim may be central to an inquiry. Relevance values must not be adjusted merely to favor a particular adequacy score; they are held fixed across the comparisons and regroupings considered below.

\section{Weighted Composition}

\label{sec:weighted-composition}

\subsection{The Contextual Valuation}

Let

\[
P=\bigwedge_{i=1}^{n}p_i
\]

\noindent satisfy the structural admissibility conditions of Section~\ref{sec:elementary}, without requiring equal relevance. For a fixed problem \(F\), write

\[
r_i=\operatorname{Rel}_F(p_i),
\qquad
R_P=\sum_{i=1}^{n}r_i.
\]

\begin{enumerate}[label=,leftmargin=0.01em,labelsep=4mm]
\item \textbf{Generalized Axiom 3} (Weighted composition). 

\textit{If \(R_P>0\), the contextual valuation is}

\[
V_{E,F}(P)
=\frac{\sum_{i=1}^{n}r_iV_E(p_i)}{R_P}.
\]
\end{enumerate}

Equivalently, the normalized weights

\[
w_i^F(P)=\frac{r_i}{R_P}
\]

\noindent are nonnegative, sum to 1, and give

\[
V_{E,F}(P)=\sum_{i=1}^{n}w_i^F(P)V_E(p_i).
\]

The notation makes explicit that the normalized weight of a component depends on the total relevance of the components included in \(P\).

If all components have the same positive relevance, this expression reduces to \mbox{Axiom 3:}

\[
V_{E,F}(P)
=\frac{1}{n}\sum_{i=1}^{n}V_E(p_i)
=V_E(P).
\]

If \(R_P=0\), the contextual valuation is undefined. This means that the conjunction supplies no content relevant to the specified assessment, not that its factual valuation is zero. Weighting does not alter the individual values \(V_E(p_i)\).

\subsection{Resolution of the Scientific Example}

\label{sec:resolution-example}

For the resistance assessment, assign

\[
\operatorname{Rel}_F(p_{0})=1,
\qquad
\operatorname{Rel}_F(p_i)=0
\quad(i=1,\ldots,n).
\]

The resistance prediction bears directly on \(F\), whereas the geometrical claims have no role in the stipulated electrical assessment. For

\[
P_n=\bigwedge_{i=0}^{n}p_i,
\]

\noindent the contextual valuation is therefore

\[
V_{E,F}(P_n)
=\frac{1(0.1)+\sum_{i=1}^{n}0\cdot1}
{1+\sum_{i=1}^{n}0}
=0.1.
\]

The result is independent of the number of irrelevant claims added. Their individual factual valuations remain unchanged, but they neither improve nor worsen the assessment of the resistance prediction.

\subsection{Consequences}

\label{sec:weighted-consequences}

The following results assume a fixed body of evidence, problem, elementary basis, and relevance assignment. All conjunctions and additions satisfy the structural admissibility conditions. Unless a zero-relevance case is explicitly considered, total relevance is \mbox{positive. Write}

\[
I_+(P)=\{i:r_i>0\}.
\]

\begin{Proposition}[Irrelevant additions]

Let \(P\) have positive total relevance, and let \(q\in\mathcal B_E\setminus S(P)\) have \(\operatorname{Rel}_F(q)=0\), and suppose that \(P\land q\) is admissible with component set \(S(P)\cup\{q\}\). Then

\[
V_{E,F}(P\land q)=V_{E,F}(P).
\]
\end{Proposition}

\begin{proof}

The additional component contributes zero to both the weighted numerator and the total relevance, whatever its factual valuation.\end{proof}

\begin{enumerate}[label=,leftmargin=0.01em,labelsep=4mm]
\item\textit{Elementary properties of the weighted mean.} 

For positive total relevance, the contextual valuation is a convex combination of the positively weighted component values. Hence,
\[
0\leq\min_{i\in I_+(P)}V_E(p_i)
\leq V_{E,F}(P)
\leq\max_{i\in I_+(P)}V_E(p_i)\leq1.
\]
It equals 1 if and only if every positively weighted component has value 1, and equals 0 if and only if all such components have value 0. The argument is the same as for the unweighted mean, applied only to positive weights. With weights fixed, increasing \(V_E(p_k)\) by \(\Delta\), within the permitted range, increases the aggregate by \(w_k^F(P)\Delta\).
\end{enumerate}

If one positively weighted component has value \(v\) and all other positively weighted components have value 1, direct substitution gives
\[
V_{E,F}(P)=1-w_k^F(P)(1-v).
\]
For \(v=0\), the loss is \(r_k/R_P\). The influence of the error therefore depends on its weight relative to the total.

For quantitative components, substituting Axiom 4 into the weighted mean gives
\[
V_{E,F}(P)=1-\frac{\sum_{i\in I_+(P)}r_i\beta_E(p_i)}{R_P}.
\]
Finally, multiplying all weights by a common positive constant leaves the ratio unchanged:
\[
\frac{\sum_i cr_iV_E(p_i)}{\sum_i cr_i}
=\frac{\sum_i r_iV_E(p_i)}{\sum_i r_i},\qquad c>0.
\]
The factor \(c\) cancels. To retain the adopted relevance scale \([0,1]\), the rescaled values must also satisfy \(cr_i\leq1\); otherwise the identity still holds for unnormalized weights.

\begin{Proposition}[Grouping invariance]

Let \(P\) and \(Q\) be admissible conjunctions with disjoint elementary component sets. Suppose their union is the component set of an admissible conjunction \(P\land Q\), with unchanged relevance assignments. If \(R_P>0\) and \(R_Q>0\), then

\[
V_{E,F}(P\land Q)
=\frac{R_PV_{E,F}(P)+R_QV_{E,F}(Q)}{R_P+R_Q}.
\]
\end{Proposition}

\begin{proof}
Write \(N_P=\sum_{p\in S(P)}\operatorname{Rel}_F(p)V_E(p)\), and define \(N_Q\) analogously. Disjointness gives the combined numerator \(N_P+N_Q\) and total relevance \(R_P+R_Q\). Substituting \(N_P=R_PV_{E,F}(P)\) and \(N_Q=R_QV_{E,F}(Q)\) yields the formula.\end{proof}

If \(R_P=0<R_Q\), then \(N_P=0\) and the combined valuation equals \(V_{E,F}(Q)\), without assigning any value to \(V_{E,F}(P)\). The corresponding statement holds with \(P\) and \(Q\) interchanged. If both totals are zero, the combined contextual valuation is undefined.

For any finite collection of disjoint groups whose conjunctions remain admissible, the pairs \((N_P,R_P)\) add componentwise. Since addition is associative, grouping does not affect the final ratio whenever the total relevance is positive. This procedure also accommodates groups of zero relevance without multiplying an undefined contextual valuation by zero.

Exact invariance under added content applies to components of zero relevance. Many components with small positive relevance can still acquire substantial total weight. The justification of the weighting scheme and the choice of elementary content therefore remain necessary parts of an application.

\section{Discussion}

\label{sec:discussion}

Partial truth makes explicit how a scientific description can retain factual adequacy despite identifiable errors. The present account separates the empirical evaluation of its components from their relevance to a specified problem. Its results concern comparisons made under stated conditions: a body of evidence, an elementary basis, discrepancy criteria, and, where applicable, relevance weights. These conditions delimit the assessment without making the facts represented depend on the assessment itself.

\subsection{Interpretation and Logical Scope}

The valuation estimates how adequately a proposition represents its reference; it does not replace the classical conditions under which the proposition is true or false. A conjunction containing a false component is classically false, yet may retain substantial factual adequacy through its other components. Likewise, a contextual value of 1 concerns the content assigned positive relevance, not the classical truth of every assertion in the description. The distinction permits the identification of errors without treating all erroneous representations as equally inadequate.

Cevolani and Schurz distinguish approximate truth, truthlikeness, and \mbox{probability \cite{CevolaniSchurz2017}.} For a finite, noncontradictory conjunction of basic statements, their initial measure of approximate truth is the fraction of true components. Our unweighted valuation coincides with this measure when the same components receive their classical values of \(0\) or \(1\). They also consider extensions beyond this initial setting, including quantitative approximation; neither component averaging nor the extension to quantitative claims is therefore distinctive of the present proposal. Their analysis additionally connects approximate truth and truthlikeness with probabilistic assessment and acceptance. The present framework has a narrower objective: to aggregate empirical adequacy estimates under explicit conditions on evidence, elementary content, discrepancy scales, and relevance to a scientific problem. It neither supplies a probabilistic acceptance rule nor measures distance from a complete truth about a domain.

Statistical methods may enter the assessment of evidence without turning the resulting adequacy score into a probability of truth. The mathematical distinction from a probability measure was established in Section~\ref{sec:immediate-consequences}.

The absence of general rules for negation and disjunction limits the scope of the theory. Such rules would be required for a recursive valuation of arbitrary formulas, but the present domain is not assumed to be closed under these connectives. Expressions involving them remain evaluable when they express propositions already admitted to the domain. Their valuation then depends on empirical criteria and elementary content, rather than on numerical operations associated with the connectives alone.

An extension would have to specify how exclusions and alternatives are empirically evaluated and what information about their contents and relations must be retained. In particular, a complement rule cannot be inferred merely from the use of the interval \([0,1]\). Proposed rules would need to respect classical equivalences, including double negation, De Morgan's laws, and absorption, wherever the expressions concerned are admitted. The grouping results established here do not demonstrate the consistency of such an extension. Developing a consistent extension with a clear empirical interpretation remains an \mbox{open challenge.}

\subsection{Relevance and the Choice of Content}
\label{sec:relevance-content}

Schurz's accounts of relevance provide a complementary comparison \cite{Schurz1991,Schurz2024}. His generalized treatment uses criteria of deductive relevance and content elements to address redundant conjunctions and irrelevant disjunctive consequences in the analysis of truthlikeness. The present framework likewise requires control over the content being evaluated, but does not derive its elementary basis from those logical criteria. It assumes a specified empirical basis and admits propositions with a unique representation satisfying the stated conditions relative to that basis. Moreover, its numerical relevance weights express the comparative importance of components for a scientific problem; they are not measures of deductive relevance. The resulting weighted adequacy score thus requires empirical and contextual choices beyond the logical identification of relevant content. The proposal does not establish a general procedure for selecting those empirical and contextual choices.

Fixing the basis prevents repetition and regrouping from changing a valuation, but does not remove every dependence on representation. Dividing one component into several claims may change both their valuations and their aggregate influence; preserving the assessment requires constraints on the valuation criteria as well as the allocation of relevance, as illustrated in Section~\ref{sec:basis-example}. The uniqueness condition on \(S(P)\) is therefore a restriction on admissible representations, not a proof that every factual proposition has a privileged decomposition. Applications must justify the adopted level of analysis; a theory of changes of basis would require additional conditions.

Weighted composition also requires more than an ordering of importance. The ratios between the weights matter: assigning twice the weight to one component gives its adequacy twice the influence in the numerator. Numerical assignments must therefore have a defensible comparative interpretation. Where only a distinction between relevant and irrelevant claims can be supported, a bivalent relevance function may be preferable to unsupported numerical precision. Where several weighting schemes are defensible, the stability of the resulting comparisons should be examined across them.

Preserving the contribution of accurate components when another component receives zero does not uniquely determine an arithmetic mean. The choice made here rests on additional assumptions: component valuations must express comparable degrees of adequacy, their contributions must be additive, and compensation between them must be appropriate to the assessment. With the component set and relevance weights fixed, a given increase in one component's valuation makes the same contribution to the aggregate regardless of the levels attained by the other components. Equal relevance gives equal influence to such increases; unequal relevance adjusts that influence in proportion to the specified weights. Normalization by total relevance ensures that a description whose relevant components all have valuation \(v\) also receives valuation \(v\). These assumptions motivate the arithmetic and weighted arithmetic means as measures of average adequacy across the selected content. Neither comparability nor the appropriateness of compensation follows merely from assigning all component values to \([0,1]\).

A plausible alternative is to take the minimum valuation among the components with positive relevance. This rule assesses the least adequate relevant component and permits no compensation for it through improvements elsewhere. For example, two descriptions with equally relevant component values \((1,0.4)\) and \((0.6,0.6)\) receive arithmetic means of \(0.7\) and \(0.6\), respectively, whereas their minima are \(0.4\) and \(0.6\). The reversal reflects different assessment questions: the mean favors greater average adequacy, while the minimum favors a higher lower bound across the assessed components. The minimum may therefore be appropriate when every relevant component must meet a common adequacy threshold. The mean is adopted here because the intended assessment concerns how much adequacy the selected content retains across its components, allowing accurate claims to contribute even when others are inaccurate.

This choice limits what a high aggregate score establishes. It does not show that every indispensable condition has been satisfied; such conditions must be assessed separately whenever the scientific problem requires them. Nor does relevance weighting eliminate compensation: exact invariance under additions holds for zero relevance, while many small positive weights may acquire substantial total influence. Applications must therefore justify the component scales, relevance assignments, and compensatory interpretation. Where those assumptions cannot be supported, the component valuations should be reported without treating their arithmetic mean as a warranted overall assessment.

\subsection{Empirical Assessment and Applications}

The examples below examine normalization, elementary content, and the comparison of rival representations. A hypothetical electrical assessment connects admissibility with sensitivity to measurement ranges, scales, and relevance; an astrophysical application then illustrates the use of published observational evidence. Together, they delimit the extension to models and theories.

\subsubsection{Normalization and Sensitivity}

For quantitative propositions, Section~\ref{sec:quantitative} specifies the interpretation and limitations of the discrepancy scale. Its consequences for actual comparisons can be examined by varying defensible normalizations while holding the evidence and asserted values fixed. Qualitative and structural claims likewise require empirical criteria that the aggregation rule alone cannot supply.

A hypothetical example illustrates sensitivity to normalization. Suppose that a positive quantity has an empirical estimate \(e=10\) units and is studied over a previously specified operating interval from \(8\) to \(12\) units. Two propositions assert values of \(10.5\) and \mbox{\(11\) units}, respectively. Taking \(s=10\) expresses their discrepancies relative to the estimated magnitude and gives valuations of \(0.95\) and \(0.90\). Taking \(s=4\), the width of the operating interval, instead expresses those discrepancies relative to the specified range of operation and gives \(0.875\) and \(0.75\). The first scale is appropriate when fractional error relative to the magnitude is the intended comparison; the second is appropriate when the fraction of the operating range is the relevant calibration. These choices must be justified by the assessment purpose before inspecting which valuation they favor.

The ordering is preserved in this example, but both the values and their separation change. More generally, for a single scalar target and a common empirical estimate, changing a positive characteristic scale cannot reverse the ordering by absolute discrepancy, although truncation can introduce ties. This limited robustness does not extend automatically to aggregates over several properties: changing their relative scales may alter their comparative contributions and hence the ranking of representations. Numerical comparisons should therefore identify the normalization used and distinguish conclusions that survive defensible alternatives from those that depend on a particular calibration.

\subsubsection{Elementary Basis and Granularity}
\label{sec:basis-example}

A hypothetical electrical assessment illustrates how the admissibility conditions can be checked. Consider two separately adjustable resistive components, with resistances \(R_1\) and \(R_2\) evaluated at a fixed temperature and under specified measurement conditions. The background assumptions identify the components, units, and experimental conditions, require positive resistances, and impose no equation determining either resistance from the other. Suppose that documented measurements yield central estimates \(e_1=100\,\Omega\) and \(e_2=200\,\Omega\), with measurement intervals specified below in Section~\ref{sec:rival-comparison}. Consider \mbox{the description}
\[
P=p_1\land p_2,\qquad
p_1:\ R_1=90\,\Omega,\qquad
p_2:\ R_2=240\,\Omega.
\]
For this assessment, choose the elementary basis \(\mathcal{B}_E=\{p_1,p_2\}\). Each component addresses a distinct empirical target. They are jointly consistent under the background assumptions: both asserted values can hold together. Neither follows from the other, since either resistance can vary while the other remains fixed. Their conjunction therefore has the unique representation \(S(P)=\{p_1,p_2\}\) within this basis; neither singleton expresses the same content. This establishes admissibility relative to the specified basis, not a uniquely privileged basis for the investigation. Joint consistency here concerns the claims and background assumptions, not exact agreement with the measured central values.

Take \(s_1=100\,\Omega\) and \(s_2=200\,\Omega\), so that discrepancies express fractional departures from the respective empirical estimates. Suppose that the assessment gives equal importance to the two targets and treats equal fractional discrepancies as comparable losses of adequacy. Then
\[
V_E(p_1)=0.9,\qquad V_E(p_2)=0.8,\qquad
V_E(P)=\frac{0.9+0.8}{2}=0.85.
\]
Reordering or repeating these claims does not change \(S(P)\) or its valuation. With more components, the grouping results likewise apply when group sizes or total weights are retained. Logical independence of the claims does not imply statistical independence of their measurements; any shared calibration limitations must still be considered in assessing the evidence.

A change of elementary basis is a different operation. Replace \(p_2\) by
\[
q_-:\ R_2\geq240\,\Omega,\qquad
q_+:\ R_2\leq240\,\Omega.
\]
Under the same interpretation of resistance, \(q_-\land q_+\equiv p_2\). The alternative basis \(\mathcal{B}'_E=\{p_1,q_-,q_+\}\) treats the lower and upper bounds as separate empirical claims. Its three members are jointly consistent, and none follows from the other two under the specified background assumptions. For example, with \(p_1\) fixed, a resistance below \(240\,\Omega\) satisfies \(q_+\) but not \(q_-\), whereas a resistance above it satisfies \(q_-\) but not \(q_+\). The bounds concern distinct exclusions, although they are assessed using the same estimate of \(R_2\).

Evaluating these bound claims requires an explicit additional discrepancy criterion. For this illustration, use the distance from the empirical estimate to the set of values allowed by each bound:
\[
d_E(q_-)=\max\{0,240\,\Omega-e_2\},\qquad
d_E(q_+)=\max\{0,e_2-240\,\Omega\}.
\]
Write \(V'_E\) for the valuation under the alternative basis and criteria. Using the same scale \(s_2\), truncation, and linear complement gives \(V'_E(q_-)=0.8\) and \(V'_E(q_+)=1\), while \(V'_E(p_1)=0.9\). If all three components in the alternative basis are assigned equal relevance, the \mbox{aggregate is}
\[
V'_E(P)=\frac{0.9+0.8+1}{3}=0.9.
\]
Thus, logically equivalent descriptions can receive different aggregate values when evaluated under different elementary bases and component assignments. No grouping result is violated: the basis and the distribution of relevance have changed.

A purely representational subdivision should be distinguished from a change in what the assessment counts as an elementary contribution. Suppose that a component with relevance \(r>0\) and valuation \(v\) is replaced by components with nonnegative relevance weights \(r_j\) and valuations \(v_j\), while all other components and their assignments remain fixed. Provided that the resulting representation satisfies the admissibility conditions, the sufficient conditions
\[
\sum_j r_j=r,\qquad
\sum_j r_jv_j=rv
\]
preserve both the denominator and numerator of the weighted mean, and therefore preserve the aggregate valuation. These equalities must be supported by the interpretation of the subdivision; they are not instructions to adjust empirical scores to obtain a preferred result.

The preceding equality-to-bounds example does not satisfy this preservation requirement under the stated equal-weight assignments. It retains the same logical content and the same empirical estimate, but treats the two bounds as separately scored contributions. It therefore changes the assessment specification without adding empirical information. If the purpose is merely to restate the original equality, its original elementary component should be retained. If the scientific question instead requires separate assessment of the lower and upper exclusions, that choice must be justified and the resulting score identified as an assessment under different criteria. Merely dividing the original relevance equally between the two bounds would not restore the original valuation: their average score would be \(0.9\), rather than the equality's \(0.8\). Preservation requires attention to both total relevance and weighted contribution.

A substantive increase in empirical resolution is different again: assessing distinct operating conditions, locations, or times may introduce claims not contained in the original description. In that case, a changed aggregate may be appropriate, but it concerns an expanded or altered empirical target. Neither logical equivalence alone nor the grouping results establish invariance across these changes.

Finally, including both the equality \(p_2\) and its two bound claims in a proposed basis \(\{p_1,p_2,q_-,q_+\}\) would give \(P\) two distinct candidate decompositions, \(\{p_1,p_2\}\) and \(\{p_1,q_-,q_+\}\), each satisfying the independence and consistency conditions. The uniqueness requirement would then fail, and the aggregate valuation would have to be withheld until the level of analysis was fixed. In applications, the empirical targets, background assumptions, and adopted granularity must therefore be specified before \mbox{comparing representations.}

\subsubsection{Rival Representations and Measurement Indeterminacy}
\label{sec:rival-comparison}

The same electrical example can illustrate the comparison of rival representations. Let \(P_A\) be the preceding description, which assigns resistances of \(90\,\Omega\) and \(240\,\Omega\), and let \(P_B\) assign \(98\,\Omega\) and \(250\,\Omega\) to the same two components under the same conditions. Extend the original equality-claim basis to include these four elementary claims. Each description selects one claim for each resistance and has a unique, jointly consistent, independent decomposition under the stated background assumptions. Both descriptions cover the same empirical targets; neither receives an advantage through omission.

Using the central estimates \(100\,\Omega\) and \(200\,\Omega\), the scales \(s_1=100\,\Omega\) and \(s_2=200\,\Omega\), and equal relevance, the component valuations are
\[
P_A:\ (0.90,0.80),\qquad
P_B:\ (0.98,0.75).
\]
Consequently,
\[
V_E(P_A)=0.85,\qquad V_E(P_B)=0.865.
\]
The second representation has greater average adequacy under these criteria, although it performs less well on the second resistance. Its advantage therefore depends on the permitted compensation between the two targets.

To examine the role of measurement indeterminacy as defined in Section~\ref{sec:quantitative}, suppose that the hypothetical measurement reports are
\[
R_1=(100\pm2)\,\Omega,\qquad
R_2=(200\pm5)\,\Omega.
\]
For this illustration, take the resulting intervals as stipulated admissible ranges after accounting for measurement and calibration limitations; no coverage probability is assigned. The comparison can be assessed by varying the reference values jointly over these ranges, while holding the asserted values, discrepancy scales, and relevance weights fixed. The same reference pair must be used for both representations.

Throughout the rectangle
\[
[98,102]\,\Omega\times[195,205]\,\Omega,
\]
the advantage of \(P_B\) remains \(0.015\). For the first resistance, its normalized discrepancy is always \(0.08\) smaller; for the second, it is always \(0.05\) larger. Equal weighting therefore gives
\[
V_E(P_B)-V_E(P_A)
=\frac{0.08-0.05}{2}=0.015.
\]
The ordering is thus robust over the stipulated measurement ranges. This is a conditional sensitivity result, not a probability that one representation is true. Considering the full rectangle does not assume statistical independence; known dependencies could restrict the admissible reference pairs further.

Now suppose that the first determination were less precise, with the same central estimate but an interval of \((100\pm5)\,\Omega\). The central scores would remain unchanged. However, at the admissible reference pair \((95,200)\,\Omega\), the scores would be \(0.875\) for \(P_A\) and \(0.86\) for \(P_B\), reversing the ordering. Under this broader measurement indeterminacy, the comparison would not be resolved by the stipulated evidence. The central scores could still be reported, but their difference would not warrant an unqualified ranking.

The comparison also depends on calibration and relevance. Suppose that the investigation had independently specified operating intervals of \([50,250]\,\Omega\) for \(R_1\) and \([100,300]\,\Omega\) for \(R_2\). Normalizing by their widths gives \(s_1=s_2=200\,\Omega\), expressing discrepancies as fractions of the operating ranges rather than of the estimated magnitudes. At the original central estimates and with equal relevance, the scores become \(0.875\) for \(P_A\) and \(0.87\) for \(P_B\). The reversal shows that robustness to measurement indeterminacy does not imply robustness to a different defensible calibration.

Finally, retaining the original scales but assigning normalized relevance \(w\) to the first target and \(1-w\) to the second gives
\[
V_{E,F}(P_B)-V_{E,F}(P_A)
=0.08w-0.05(1-w)=0.13w-0.05.
\]
Thus, \(P_B\) ranks higher when \(w>5/13\), the descriptions tie when \(w=5/13\), and \(P_A\) ranks higher when \(w<5/13\).

A high average must also be distinguished from satisfying an indispensable requirement. Suppose that the electrical application independently requires the prediction of the second resistance to differ from the empirical reference by no more than \(20\,\Omega\). Throughout the stipulated reference interval \([195,205]\,\Omega\), the discrepancies of \(P_A\) and \(P_B\) on this component exceed that tolerance. Both descriptions therefore fail this requirement despite their relatively high central aggregate scores. The component discrepancy and the failure must be reported separately; changing relevance weights does not establish compliance. This acceptance threshold is distinct from the discrepancy scale used to calculate graded adequacy. If the application also concerns a third property for which no adequate evidence is available, that claim remains unevaluated and must be reported as outside the assessed coverage. It receives neither a zero adequacy score nor an assignment of zero relevance merely because evidence is lacking.

The comparison consequently supports a limited conclusion about scientific improvement. With the original scales, equal relevance, and the narrower measurement ranges, \(P_B\) has greater average factual adequacy than \(P_A\) across the two assessed targets. It nevertheless performs less well on the second target, and neither description meets the additional requirement just specified. Moreover, the ordering changes under alternative calibration or relevance assignments. The example thus illustrates a conditional improvement in average adequacy, not an unqualified advance in every scientifically important respect. Such improvement can contribute to scientific progress when assessed alongside coverage, explanatory scope, and indispensable requirements, but the aggregate score alone does not establish progress independently of those considerations.

\subsubsection{An Astrophysical Application}

An illustrative, concrete application to a scientific problem concerns the mass-loss rate of a massive star. Massive stars expel matter through stellar winds, and the rate at which they lose mass is important for understanding both their evolution and their interaction with the surrounding interstellar medium. A quantitative statement about this rate can be assessed by comparing its asserted value with an estimate inferred from observations of the wind.

Consider the massive star \(\zeta\) Puppis. Cohen et al.~\cite{Cohen2010} analyzed 16 X-ray emission lines or line complexes observed with \textit{Chandra} space telescope on  28--29 March 2000. By modeling their profiles, including absorption by the stellar wind, they inferred a mass-loss rate with central estimate
\[
e_E=3.5\times10^{-6}\,M_\odot\,\mathrm{yr}^{-1}
\]
\textls[-20]{and a reported statistical contribution to measurement indeterminacy of \(\pm0.3\times10^{-6}\,M_\odot\,\mathrm{yr}^{-1}\).} The authors also identify substantial systematic limitations associated with the elemental abundances adopted in calculating the wind opacity. They estimate this systematic contribution to the indeterminacy of the inferred mass-loss rate to be a factor of two, substantially larger than the quoted statistical contribution. 
This determination therefore provides an empirical estimate under specified modeling assumptions, rather than a direct measurement independent of those assumptions.

Consider, for illustration, the proposition \(p\) asserting that the mass-loss rate characterizing the wind during this observation was
\[
p:\quad \dot M=4.0\times10^{-6}\,M_\odot\,\mathrm{yr}^{-1}.
\]
The asserted value is introduced here to illustrate the assessment and is not attributed to Cohen et al. Let \(E\) comprise the statements reporting these observations, together with the documentation supporting the inferred estimate. Taking its central estimate as the characteristic scale, \(s_E(\dot M)=|e_E|\), expresses the discrepancy as a fraction of the empirically inferred rate. Treating \(p\) as an elementary quantitative claim at the adopted level of analysis, Axiom 4 gives
\[
V_E(p)
=1-\min\left\{1,
\frac{|4.0-3.5|}{3.5}
\right\}
=\frac{6}{7}\approx0.857.
\]

The common units cancel in the ratio. Under this normalization, the proposition receives a substantial estimated degree of factual adequacy despite its discrepancy from the central estimate. The result is neither a probability that \(p\) is true nor a measure of confidence in the inferred rate. The statistical interval and systematic limitations must accompany the assessment; the score does not establish that the asserted and inferred rates are empirically distinguishable. A revision of the empirical estimate or its supporting assumptions may warrant a revised valuation.

\subsubsection{Model Assessment and Coverage}

If a stellar-wind model yields such a proposition as a prediction, the calculation assesses that particular consequence within the specified observational domain. Extending the assessment to models and theories requires selecting finite sets of testable consequences that satisfy the admissibility conditions, with appropriate discrepancy scales and justified relevance assignments. Their valuation concerns those consequences within the tested domain, not every assertion or implication of the theory. Comparisons between rival representations require corresponding empirical targets, compatible discrepancy scales, and common relevance criteria. Unexamined claims cannot be silently treated as either accurate or inaccurate.

Adequacy and coverage must also be distinguished. A representation may receive a high value for the few claims it makes while leaving much of the scientific question unanswered. The present normalized mean does not automatically penalize omissions. Applications should therefore report what was assessed and what remains outside the assessment, together with the component values that the aggregate summarizes. Progress can then be examined as improvement in factual adequacy under comparable conditions, alongside changes in explanatory scope and empirical coverage, without interpreting any finite score as a guarantee of complete truth.

\section{Conclusions}
\label{sec:conclusions}

The factual adequacy of a scientific proposition cannot be assessed by logical analysis alone. It requires comparing what is asserted with evidence concerning the facts represented. The valuation function

\[
V_{E}:\mathcal{D}_{E} \rightarrow [0,1]
\]

\noindent expresses the resulting estimate relative to a body of evidence \(E\) and specified valuation criteria. Its target is adequacy to the facts for fixed content and valuation criteria; its values represent neither degrees of confidence nor probabilities. The passage from agreement with an empirical estimate to estimated factual adequacy depends on the methodological warrant for that estimate. Correcting a systematic bias can therefore require revising a valuation, while an unresolved defect in the evidence may require withholding it.

For admissible finite conjunctions, the present formulation aggregates the valuations of distinct elementary components within a fixed basis. The arithmetic mean applies to equal relevance, while the weighted mean accommodates differences in relevance. A component assigned zero does not cancel the adequacy contributed by other components with positive weight. Evaluating the complete component set, or combining disjoint groups while retaining their total weights, makes the result independent of grouping. This removes the dependence on parenthesization discussed in \cite{Gracia2025} without treating the binary arithmetic mean as an associative operation.

These results concern estimated adequacy, not a replacement for classical logic. Logical equivalence constrains the identification of the elementary content being evaluated; the composition rules do not define a recursive calculus for arbitrary formulas. No general numerical rules for negation or disjunction are postulated.

For quantitative components, the aggregate is one minus their mean normalized discrepancy, weighted by relevance where applicable. Its endpoints concern the adopted adequacy scale, not the classical truth conditions of conjunction or complete correspondence established independently of empirical limitations. The worked comparison shows why central scores must be accompanied by an assessment of measurement indeterminacy and sensitivity to scales and relevance weights.

Truth and relevance remain distinct. A proposition may be accurate without contributing to the problem being addressed, or relevant despite being inaccurate. Holding the elementary basis and relevance assignments fixed, adding admissible components with zero relevance leaves the contextual valuation unchanged. This protection does not extend to components with small positive relevance, whose weights may accumulate. Because the weighted mean permits compensation between components, an important error can still coexist with a high aggregate score. Indispensable conditions must therefore be \mbox{checked separately.}

The framework supplies aggregation rules, not a universal method for choosing elementary components, discrepancy scales, or relevance weights. Applications must justify those choices and examine the stability of comparisons when defensible alternatives exist. Further work is needed on changes of elementary basis, empirical criteria for qualitative claims, and the joint assessment of correlated or conflicting evidence. For models and theories, valuations concern selected finite sets of testable consequences satisfying the admissibility conditions, not every assertion or implication of the theory. Adequacy must also be distinguished from coverage: a high score does not show that all relevant questions have been answered.

The distinction between factual adequacy, confidence, and relevance can also inform everyday judgment, although ordinary evidence seldom meets the conditions imposed here on scientific bodies of evidence. Perceptions, memories, testimony, and documents may support informal assessments without numerical truth values or relevance weights. A degree of confidence that things are as asserted must still be distinguished from an assessment of how closely an assertion represents them. The required scrutiny of evidence, assumptions, and limitations should reflect the consequences of error.

Science makes these demands systematic by testing representations, correcting their errors, and identifying the respects in which they remain adequate. Progress can be assessed through improvements in factual adequacy under comparable conditions, alongside gains in explanatory scope and empirical coverage, without assuming demonstrable convergence toward a final, complete truth. A theory of partial truth does not replace the search for truth; it clarifies how a fallible representation can retain factual adequacy and how that adequacy can be assessed.

\vspace{6pt}

\funding{This research received no external funding.}

\institutionalreview{Not applicable.}

\informedconsent{Not applicable.}

\dataavailability{No new data were created or analyzed in this study. Data sharing is not applicable to this article.} 

\conflictsofinterest{The author declares no conflicts of interest.}

\begin{adjustwidth}{-\extralength}{0cm}

\printendnotes[custom]

\end{adjustwidth}
\reftitle{References}

\begin{adjustwidth}{-\extralength}{0cm}

\PublishersNote{}
\end{adjustwidth}
\end{document}